\documentstyle[twocolumn,prl,aps]{revtex}
\newcommand{\be}{\begin{equation}} \newcommand{\ee}{\end{equation}} 
\newcommand{\bea}{\begin{eqnarray}}\newcommand{\eea}{\end{eqnarray}}
\newcommand{\bm}[1]{\mbox{\boldmath$#1$}}
\newcommand{\grad}{\bm \nabla}
\begin{document}
\draft
\title{ Conformal symmetry and the nonlinear Schr$\ddot{o}$dinger equation}
\author{Pijush K. Ghosh$^{*}$}
\address{ 
Department of Physics,
Ochanomizu University,\\
2-1-1 Ohtsuka, Bunkyo-ku,
Tokyo 112-8610, Japan.\\}

\maketitle
\begin{abstract} 
We show that the width of the wave-packet of a class of generalized nonlinear
Schr$\ddot{o}$dinger equations (NLSE) trapped in an arbitrary time-dependent
harmonic well in any dimensions is universally determined by the same Hill's
equation. This class of generalized  NLSE is characterized by a dynamical
$O(2,1)$ symmetry in absence of the trap. As an application, we study the
dynamical instabilities of the rotating as well as non-rotating Bose-Einstein
condensates in one and two dimensions. We also show exact extended
parametric resonance in a non-relativistic Chern-Simons theory producing
a gauged NLSE.
\end{abstract}
\narrowtext

The nonlinear Schr$\ddot{o}$dinger equation (NLSE) appears in many branches
of present-day physics and mathematics\cite{theo}. The optical
soliton\cite{hase}, a solution of the NLSE, has even been observed
experimentally\cite{exp}. The
one dimensional NLSE is exactly solvable. There exists several other
generalized NLSE in one dimension which are also exactly solvable. However,
a very little exact and analytical results are known for higher dimensional
generalizations of these models, although they are very much relevant in many
branches of modern science.
The purpose of this letter is to present an exact, analytical description of
the dynamics of the width of the wave-packet for a class of generalized NLSE
in arbitrary dimensions trapped in a time-dependent harmonic well. This class
of generalized NLSE is characterized by a dynamical $O(2,1)$ symmetry in
absence of the trap.

Consider the following Lagrangian in arbitrary $d+1$ dimensions,
\be
{\cal{L}} = i \psi^* \partial_{\tau} \psi 
- \frac{1}{2 m} {\mid \grad \psi \mid}^2
- g V(\psi, \psi^*, {\bf r}).
\label{eq1}
\ee
\noindent The coupling constant $g$ has the inverse-mass dimension in the
natural units with $c=\bar{h}=1$. The real potential $V$ does not depend on
any dimensional coupling constant. This allows to have a scale-invariant
theory. We demand the invariance of the action 
${\cal{A}} =\int d \tau d^d{\bf r} {\cal{L}}$ under the following
time-dependent transformations\cite{trans,st},
\bea
&& {\bf r} \rightarrow {\bf r_h} = {\dot{\tau}}(t)^{-\frac{1}{2}} {\bf r}, \ \
\tau \rightarrow t = t(\tau), \ \ \dot{\tau}(t) =
\frac{d \tau(t)}{d t}, \nonumber \\
&& \psi(\tau, {\bf r}) \rightarrow \psi_h(t, {\bf r}_h) =
\dot{\tau}^{\frac{d}{4}} exp \left ( - i m \frac{\ddot{\tau}}{4 \dot{\tau}}
r_h^2 \right ) \psi(\tau, {\bf r}),
\label{eq5}
\eea
\noindent with the scale-factor $\tau$ given by,
\be
\tau(t) = \frac{ \alpha t + \beta}{\gamma t + \delta}, \ \
\alpha \delta - \beta \gamma =1.
\label{eqss}
\ee
\noindent Particular choices of $\tau(t)=  t + \beta, \alpha^2 t$,
and $ \frac{t}{1 + \gamma t}$, correspond to time translation, dilation
and special conformal transformation. The generators of these 
transformations ( as given below ) close under an $O(2,1)$ algebra.
Although $V$ is restricted to have specific forms due to the requirement of
the symmetry, one can still make infinitely many choices of it. It might
be noted here, apart from its dependence on the condensate $\psi$,
the potential can also be explicitly dependent on the 
space-coordinates. We keep $V$ arbitrary, but, consistent with the $O(2,1)$
symmetry in (\ref{eq1}), unless mentioned otherwise.

Let us now introduce two moments $I_1$ and $I_2$ in terms
of the density $\rho$ and the current ${\bf j}$ as,
\bea
&& \rho (\tau, {\bf r}) = \psi^* \psi, \ \
{\bf j}(\tau, {\bf r}) = -\frac{i}{2 m} \left ( \psi^* 
{\bf \grad} \psi - \psi { \grad}
\psi^* \right ),\nonumber \\
&& I_1 (\tau) = \frac{m}{2} \int d^d {\bf r} \ r^2 \ \rho, \ \
I_2 (\tau) = \frac{m}{2} \int d^d {\bf r} \ {\bf r} \cdot {\bf j}.
\label{eq2}
\eea
\noindent We are dealing with a conservative system and the total number
of particles $ N(\tau)= \int d^d {\bf r} \rho$ is a constant of motion.
The moment $I_2$ is related to the speed of the growth of the condensate. The
moment $I_1$ describes the square of the width of the wave-packet\cite{def}.
This quantity plays the central role in the
analysis of the collapse of the condensates of the NLSE with or without
time-independent harmonic trap \cite{za,mw,bull,spain}. It is also used in the
context of the Bose-Einstein condensation(BEC)\cite{gp} to study the low
energy excitations and in optics\cite{opt} to determine the beam parameter
evolution. The dynamics of $I_1$,
when the system (\ref{eq1}) is immersed in an external time-dependent
harmonic trap, is the central subject of the investigation of this letter.
We show that the
dynamics of $I_1$ is universally determined by the same solvable Hill's
equation, independent of the space-dimensionality, integrability and nature
( short-range, long-range, local, non-local, linear, non-linear ) of the
interaction. The universality in the description of the dynamics of the width
for this class of theory  has been observed partially through a time-dependent
variational analysis in \cite{tkg}. We present here exact, analytical and
complete treatment.

The system (\ref{eq1}) has a dynamical $O(2,1)$ symmetry. The generators,
the Hamiltonian $H$, the dilatation generator $D$ and the conformal
generator $K$ are,
\bea
&& H = \int d^d {\bf r} \left [ \frac{1}{2 m}
{\mid \grad \psi \mid}^2 + g V(\psi, \psi^* , {\bf r})
\right ],\nonumber \\
&& D =  \tau H - I_2,\nonumber \\
&& K = - \tau^2 H + 2 \tau D + I_1.
\label{eq3}
\eea
\noindent These generators are constant in time and lead to the
following equations,
\be
\frac{d H}{d \tau} = 0, \ \
\frac{d I_1}{d \tau} = 2 I_2, \ \
\frac{d I_2}{d \tau} = H.
\label{eq4}
\ee
\noindent For time independent solutions, both $I_1$ and $I_2$ do not depend
on $\tau$. As a consequence, the static solutions of a system with $O(2,1)$
symmetry carry zero
energy \cite{cz}. We also note that $I_2=D=0$ and $K=I_1$ for static solutions
of (\ref{eq1}). Defining the width of the wave-packet, $X=\sqrt{I_1}$, it is
easy to find a decoupled equation for $X$ from (\ref{eq4}),
\be
\frac{d^2 X}{d \tau^2} = \frac{Q}{X^3}, \ \
Q = I_1 H - I_2^2 > 0, \ \ \frac{d Q}{d \tau}=0.
\label{eq4.1}
\ee
\noindent The constant of motion $Q$ is the Casimir operator of the
$O(2,1)$ symmetry. Eq. (\ref{eq4.1}) can be interpreted as the equation of
motion of a particle moving in an inverse-square potential. Interestingly
enough, this system also has a dynamical $O(2,1)$ symmetry. This reduced
system of a particle in an inverse-square potential is a
well-studied problem and the solution is given by\cite{trans},
\be
X^2 = ( a + b \tau )^2 + \frac{Q}{a^2} {\tau}^2,
\label{eq4.2}
\ee
\noindent where $a$ and $b$ are the integration constants.

Consider the time-dependent transformations in (\ref{eq5}) with arbitrary
scale-factor $\tau(t)$. This is no more symmetry transformations of
(\ref{eq1}) for general $\tau$. The action $ {\cal{A}}$ is transformed into
a new one,
${\cal{A}}_h = \int dt d^d {\bf r}_h {\cal{L}}_h$, containing a time-dependent
harmonic trap. The new Lagrangian ${\cal{L}}_h$ now reads,
\bea
{\cal{L}}_h  & = & i \psi_h^* \partial_t \psi_h  
- \frac{1}{2 m} {\mid \grad_h \psi_h \mid}^2\nonumber \\
& - & g V(\psi_h, \psi_h^*, {\bf r}_h)
- \frac{1}{2} m \omega(t) r_h^2 {\mid \psi_h \mid}^2.
\label{eq6}
\eea
\noindent The time-dependent frequency $\omega(t)$ of the harmonic trap is
determined by,
\be
\ddot{b} + \omega(t) b = 0, \ \ b(t) = \dot{\tau}^{-\frac{1}{2}}.
\label{eq7}
\ee
\noindent Once the solution of the equation of motion of (\ref{eq1}) is known,
the same can be obtained for (\ref{eq6}) by using the transformation
(\ref{eq5}) and the equation (\ref{eq7}) or the vice versa. Equation
(\ref{eq7}) describes the motion of a particle in a time-dependent harmonic
trap. For $\tau(t) = \frac{1}{\omega_0} tan (\omega_0 t)$, it gives
$\omega=\omega_0$. For the special choice of (\ref{eqss}), the frequency
$\omega$ obviously vanishes. The general solution of (\ref{eq7}) for the
physically relevant periodic $\omega(t)$ is well-known and will be discussed
below.

The dynamical $O(2,1)$ symmetry of ${\cal{L}}$ is not present
for ${\cal{L}}_h$. We replace $({\bf r}, \psi, \tau)$ by $({\bf r}_h,
\psi_h, t)$ in the definition of $I _1$, $I_2$ and $H$ and denote the
resulting expressions in terms of the `curly form' of the associated
variables. Under the transformation (\ref{eq5}), the equations
(\ref{eq4}) have the following form,
\bea
&& \dot{\cal{I}}_1(t) = 2 {\cal{I}}_2(t),\nonumber \\
&& \dot{\cal{I}}_2(t) = {\cal{H}}(t) - \omega(t) {\cal{I}}_1(t),\nonumber \\
&& \dot{\cal{H}}(t) = - 2 \omega(t) {\cal{I}}_2(t).
\label{eq8}
\eea
\noindent Defining a new variable ${\cal{X}}(t) =
\sqrt{{\cal{I}}_1(t)}$, it is easy to find a decoupled equation for
${\cal{X}}$,
\bea
&& \ddot{\cal{X}} + \omega(t) {\cal{X}} = \frac{\cal{Q}}{{\cal{X}}^3},
\nonumber \\
&& {\cal{Q}} = {\cal{I}}_1 {\cal{H}} - {\cal{I}}_2^2 > 0, 
\ \ \dot{\cal{Q}}=0.
\label{eq9}
\eea
\noindent We have the surprising result that the dynamics of the width
of the wave-packet of the system (\ref{eq6}) is universally determined by
the equation (\ref{eq9}). This result is independent of the integrability
of the model. We also have the freedom of choosing a large class of $V$
as long as the dynamical $O(2,1)$ symmetry in absence of the harmonic trap
is maintained. The knowledge of the time-evolution of ${\cal{X}}$ allows us to
determine the time-evolution of ${\cal{H}}$,
\be
{\cal{H}} = \dot{\cal{X}}^2 + \frac{\cal{Q}}{{\cal{X}}^2}.
\ee
\noindent For the time-independent trap, $\omega(t)=\omega_0$, the Hamiltonian
$H_h = {\cal{H}} + \omega_0 {\cal{I}}_1$ corresponding to the Lagrangian
${\cal{L}}_h$ is a constant of motion. The Hamiltonian $H_h$ is related to
the generator of the compact $SO(2)$ rotation of $SO(2,1)$.

Eq. (\ref{eq9}) can be interpreted as that of a particle moving in a
time-dependent harmonic trap and a inverse-square potential. Due to the
underlying $O(2,1)$ symmetry in absence of the trap, this equation
can be obtained directly from (\ref{eq4.1}) through the use of the
transformations (\ref{eq5}). The dynamics of the width ${\cal{X}}$ can thus be
constructed exactly from (\ref{eq4.2}),
\be
{\cal{X}}(t) = b(t)  X(\tau(t)),
\ee
\noindent with the knowledge of the scale-factors $\tau(t)$ and $b(t)$
from (\ref{eq7}) for a particular choice of $\omega(t)$. We provide below
a familiar form of solution
of Eq. (\ref{eq9}), since it appears in many branches of physics including the
cylindrically symmetric two-dimensional NLSE\cite{spain}. The general solution
of (\ref{eq9}) is given by,
\be
{\cal{X}}^2(t) = u^2(t) + \frac{Q}{W^2} v^2(t), \ \ 
W(t) = u \dot{v} - v {\dot{u}},
\ee
\noindent where $u(t)$ and $v(t)$ are two independent solutions
of the equation,
\be
\ddot{x} + \omega(t) x =0,
\ee
\noindent satisfying $u(t_0)={\cal{X}}(t_0), \dot{u}(t_0) = \dot{\cal{X}}(t_0),
\dot{v}(t_0)=0$ and $v(t_0) \neq 0$. For periodic $\omega(t)$ with
the period $T$, the above equation is known as the Hill's equation
and is a text-book material\cite{arnold}. We just mention here
the general stability criteria in terms of the quantity
$\delta = {\mid u(T) + \dot{v}(T) \mid}$ with the normalization
${\cal{X}}(0)=0, \dot{\cal{X}}(0)=1$ and $v(0)=1$.
The solution is stable for $\delta < 2$, while it is unstable for
$\delta > 2$. We remark that the same stability criteria is valid for
Eq. (\ref{eq7}).

The central result of this letter is contained in (\ref{eq9}).
Although our main concern in this letter is on NLSE, we remark
that the same result is true for a class of linear Schr$\ddot{o}$dinger
equations with Calogero-type inverse-square interaction in arbitrary
dimension\cite{me1}. An example of arbitrary $d$ dimensional non-linear
potential consistent with $O(2,1)$ symmetry is given by,
\be
V(\psi, \psi^*, {\bf r}) = \int d^d {\bf r}^{\prime}
\psi^*({\bf r}^{\prime})
U({\bf r} - {\bf r}^{\prime}) \psi({{\bf r}}^{\prime})
{\mid \psi({\bf r}) \mid}^2,
\ee
\noindent with $U(r)$ having the following scaling property.
For $ {\bf r} \rightarrow \epsilon {\bf r}, \
U({\bf r}) \rightarrow U(\epsilon {\bf r}) = \epsilon^{-2} U({\bf r})$.
We now discuss a few specific examples of NLSE with different choices of
$V$ which are relevant in the contemporary literature.

{\it BEC in} $ d=1, g > 0, V={\mid \psi \mid}^6$ :
The Gross-Pitaevskii equation (GPE) describing the repulsive Bose-Einstein
condensates trapped in a time-dependent harmonic trap in one
dimension can be obtained from the Lagrangian (\ref{eq6}) \cite{kolo}. 
Exact soliton solutions of the GPE equation have been obtained in absence
of the trap \cite{kolo,mumu}. Only approximate or numerical results are known,
when the time-independent harmonic trap is included\cite{kolo}.

Following our analysis, the exact solutions of (\ref{eq6}) can be obtained
from those of (\ref{eq1}) by
simply using the transformation (\ref{eq5}) and the equation (\ref{eq7})
determining the time-dependent scale-factor $\tau$ for a particular choice
of the $\omega(t)$. We consider the case of time-independent trap
with $\omega(t)=\omega_0$ and choose $g=\frac{\pi^2}{6 m}$. Define the
following dimensionless variables,
\be
\bar{x}_h = \pi \psi_0^2 x_h, \ \
\bar{t} = \frac{\pi^2 \psi_0^4}{m} t, \ \
\bar{\psi}_h = \frac{\psi_h} {\psi_0}, \ \
\bar{\omega}_0 = \frac{m \omega_0}{\pi^4 \phi_0^8},
\ee
\noindent where $ \psi_0^2$, the asymptotic value of the density, is related
to the chemical potential $\mu$ by, $\psi_0^2 = \frac{\sqrt{2 m \mu}}{\pi}$.
The exact solution for $\bar{\psi}_h$ is,
\bea
\bar{\psi}_h  & = & \frac{1}{cos(\bar{\omega}_0 \bar{t})}
exp \left ( - \frac{i \bar{\omega}_0}{2} tan(\bar{\omega}_0 \bar{t})
\bar{x}_h^2
-\frac{i}{\bar{\omega}_0} tan(\bar{\omega}_0 \bar{t}) \right )\nonumber \\
&& \times \left [ \frac{cosh[\frac{2 y}{cos(\bar{\omega}_0 \bar{t})}] 
- 1}{cosh[\frac{2 y}{cos(\bar{\omega}_0 \bar{t})}] +2} \right ]^{\frac{1}{2}}.
\eea
\noindent In the limit $\bar{\omega}_0 \rightarrow 0$, the solution for the
system without the trap is recovered\cite{kolo}. Without loss of any
generality, we are choosing $\beta=0$ in Eq. (12) of \cite{kolo}.

{\it BEC in} $d=2, V={\mid \psi \mid}^4$:
We get the GPE describing two dimensional BEC in a time-dependent trap from
the Lagrangian (\ref{eq6}). No exact solution of this GPE with or without
the trap is known. However, this is a well-studied system and many of the
dynamical properties are already known\cite{kagan,rp,spain}.
We concentrate here on the rotating BEC and present some new interesting
results. Consider a further time-dependent rotation in (\ref{eq6})\cite{st},
\be
t \rightarrow \tilde{t}=t, \ \
{\bf r}_h \rightarrow \tilde{{\bf r}}_h = \left ( \matrix{ {cos f(t)} &
{sin f(t)} \cr\\ {-sin f(t)} & {cos f(t)}} \right ) {\bf r}_h.
\label{eq15}
\ee
\noindent This transforms ${\cal{A}}_h$ to
$\tilde{{\cal{A}}}_h=\int d\tilde{t} d^d \tilde{\bf r}_h \tilde{\cal{L}}_h$
containing an additional term proportional to the $z$ component of the angular
momentum with the coefficient given by a time-dependent frequency. In
particular, the new $\tilde{\cal{L}}_h$ is given by,
\bea
\tilde{\cal{L}}_h  & = & i \psi_h^* \partial_{\tilde{t}} \psi_h  
- \frac{1}{2 m} {\mid \tilde{\grad}_h \psi_h \mid}^2
- g V(\psi_h, \psi_h^*, \tilde{{\bf r}}_h)\nonumber \\
& - & \frac{1}{2} m \omega(t) \tilde{r}_h^2 {\mid \psi_h \mid}^2
- \dot{f} \ \psi_h^* L_z \psi_h,\nonumber \\
L_z & = & - i \left ( \tilde{x}_h \frac{\partial}{\partial \tilde{y}_h} - 
\tilde{y}_h \frac{\partial}{\partial \tilde{x}_h} \right ),
\label{eq16}
\eea
\noindent where $\tilde{x}_h$ and $\tilde{y}_h$ are the components of the
two dimensional vector $\tilde{{\bf r}}_h$. This is the Lagrangian
for rotating BEC in an external time-dependent isotropic trap in two
dimensions\cite{rotate}. Interestingly, once the solution of the equation of
motion of (\ref{eq1}) is known, the same can be obtained for (\ref{eq16}),
using the equations (\ref{eq5}), (\ref{eq7}) and (\ref{eq15}), or vice versa.
Moreover, under the transformation (\ref{eq15}), the set of equations in
(\ref{eq8}) remains the same in terms of the new variables $\tilde{t}$ and
$\tilde{\bf r}_h$. Thus, the dynamics of the width ${\cal{X}}({\tilde{t}})$
of (\ref{eq16}) is again universally determined by the equation (\ref{eq9}).
The introduction of the last term in (\ref{eq16}) does not change the
dynamical properties of the width.
There may be dynamic instabilities solely
due to the rapid fluctuations in
the phase of the condensate during the evolution in time. However,
it is obvious from the definition of $I_1$ that such instabilities do not show
up in the evolution of the width.\\
{\it Gauged NLSE}: We now show that the dynamics of the width remains
unchanged even if the non-trivial gauge-fields are introduced in (\ref{eq1})
maintaining the $O(2,1)$ symmetry. Consider a Lagrangian in $2+1$ dimensions
with the gauge-fields $(A_0, {\bf A})$ and the matter-field $\psi$,
\bea
{\cal{L}}_g & = & i \psi^* (\partial_{\tau} - i A_0) \psi - \frac{1}{2 m}
{\mid (\grad - i {\bf A}) \psi \mid}^2\nonumber \\
& - & g V(\psi, \psi^*, {\bf r})
+ \frac{\kappa}{4} \epsilon^{\mu \nu \lambda} F_{\mu \nu} A_{\lambda},
\eea 
\noindent where the last term is the Chern-Simons term. The moment $I_1$
for this non-relativistic Chern-Simons(CS) theory can be interpreted as the
width of the soliton or alternatively as the quadrupole moment. For
$V=\frac{1}{2} {\mid \psi \mid}^4$, this is the Jackiw-Pi model describing
gauged NLSE \cite{jp}. This is relevant in theories with anyons and in the
quantum Hall effect\cite{jp,ez,me}. The Jackiw-Pi model is exactly solvable
at the self-dual point, $ g = \frac{1}{m {\mid \kappa \mid}}$. Our
result is valid
for arbitrary $V$ maintaining $O(2,1)$ symmetry. For the particular case
of Jackiw-Pi model, the importance of our result lies at all non-self-dual
points, where the model is not integrable. The Hamiltonian is given by,
\be
H = \int d^2 {\bf r} \left [ \frac{1}{2 m} {\mid (\grad - 
i {\bf A}) \psi \mid}^2 + g V(\psi, \psi^*, {\bf r}) \right ].
\label{nh}
\ee
\noindent The CS term being a topological term do not contribute
to the Hamiltonian. 
The generators $D$ and $K$ have the same expressions as in (\ref{eq3}),
with the partial derivative in the expression of the current ${\bf j}$ in
the definition of the moment $I_2$ replaced by the respective
covariant derivative. The treatment is now identical to the case
without the gauge-fields. The same transformations (\ref{eq5})
with $d=2$ and the gauge-fields transforming accordingly,
\bea
&& A_{\mu}^h (t, {\bf r}_h) = \frac{\partial x^{\nu}}{\partial x_h^{\mu}}
A_{\nu}(\tau, {\bf r}),\nonumber \\
&& x^{\mu}=(\tau, {\bf r}),\ \ A^{\mu} =(A_0, {\bf A}), \ \ \mu=0, 1,2,
\eea
\noindent introduce a time-dependent harmonic trap\cite{jp,ez,me}.
The width of the soliton of this new Lagrangian is again universally
determined by (\ref{eq9}).
Interestingly, the introduction of the gauge-fields and the nontrivial
CS term to the usual two dimensional NLSE does not change the
dynamics of ${\cal{X}}$. A comment is in order at this point. It is known
that Eq.(\ref{eq9}) admits parametric resonances. Thus, the solitons of
the non-relativistic CS theory should exhibit the same phenomenon.
This provides an example of exact, extended parametric resonance in a gauge
theory with the nontrivial CS term.

In conclusion, we have shown that the width of the wave-packet of a class
of generalized NLSE is universally determined by the same Hill's equation.
This class of NLSE is characterized by a dynamical $O(2,1)$ symmetry in absence
of the trap.
The result is so robust that it is independent of, (i) the space
dimensionality, (ii) the integrability of the model, and (iii) short-range,
long-range, local, non-local, linear or non-linear nature of the many-body
interaction. This result persists with its full generality even when the
gauge-fields are introduced maintaining the dynamical $O(2,1)$ symmetry. The
later example allows us to study an exact parametric resonance in a theory
with the nontrivial gauge fields. Special cases of this class of generalized
NLSE
are relevant in BEC and in non-relativistic Chern-Simons theory. It would be
nice to see the importance of this class of NLSE in many more physical systems.
\acknowledgements{This work is supported by a fellowship (P99231) of the 
JSPS. I would like to thank T. Deguchi and T. K. Ghosh for their continuous
interest in this work and valuable comments on the manuscript. }



\begin{thebibliography}{99}

\bibitem{theo} V. E. Zakharov and A. B. Shabat, Sov. Phys. JETP {\bf 34},
62 (1972); R. Camassa, J. M. Hyman and B. P. Luce, Physica {\bf 123D},
1 (1998).

\bibitem{hase} A. Hasegawa and F. Tappert, Appl. Phys. Lett. {\bf 23},
142 (1973); {\bf 23}, 171 (1973).

\bibitem{exp} L. F. Mollenauer, R. H. Stolen and J. P. Gordan, Phys.
Rev. Lett. {\bf 45}, 1095 (1980).

\bibitem{trans} V. de Alfaro, S. Fubini and G. Furlan, Nuvo Cimento {\bf A34},
569 (1976); R. Jackiw, Ann. Phys. (N.Y.) {\bf 129}, 183(1980); {\bf 201}, 83
(1990).

\bibitem{st} S. Takagi,
Prog. Theor. Phys. {\bf 84}, 1019 (1990); {\bf 85}, 463 (1991); {\bf 85},
723 (1991).

\bibitem{def} We are using the phrase `width of the wave-packet' following
Ref. \cite{spain}. It may be more appropriate for some specific cases to
identify $I_1$ with a different physical quantity instead of the width.
The central result, the exact and the universal description of the dynamics
of the moment $I_1$, is independent of any such change in the nomenclature.

\bibitem{za} V. E. Zakharov, Sov. Phys. JETP {\bf 35}, 908 (1972);
V. E. Zakharov and V. S. Synakh, Sov. Phys. JETP {\bf 41}, 465 (1975);
M. I. Weinstein, Commun. Math. Phys. {\bf 87}, 567 (1983).

\bibitem{mw} T. Tsurumi and M. Wadati, J. Phys. Soc. Japan {\bf 66},
3031 (1997); {\it ibid} {\bf 66}, 3035 (1997); M. Wadati and T. Tsurumi,
Phys. Lett. {\bf A 247}, 287 (1998); T. Tsurumi, H. Morise and M. Wadati,
Int. Jour. Mod. Phys. {\bf B 14}, 655 (2000), cond-mat/9912470.

\bibitem{bull} A. V. Rybin, G. G. Varzugin, M. Lindberg, J. Timonen
and R. K. Bullough, Phys. Rev. {\bf E 62}, 6224 (2000),
cond-mat/0001059.

\bibitem{spain} J. J. Garcia-Ripoll and V. M. Perez-Garcia,
Phys. Rev. Lett. {\bf 83}, 1715 (1999)
; patt-sol/9904006.

\bibitem{gp} S. Stringari, Phys. Rev. Lett. {\bf 77}, 2360 (1996);
V. M. Perez-Garcia et. al., Phys. Rev. Lett. {\bf 77}, 5320 (1996);
F. Dalfovo, S. Giorgini, L. P. Pitaevskii and S. Stringari,
Rev. Mod. Phys. {\bf 71} 463 (1999), cond-mat/9806038.

\bibitem{opt} S. N. Vlasov, V. A. Petrischev and V. I. Talanov, Radiophys.
Quantum Electron. {\bf 14}, 1062 (1971); V. M. Perez-Garcia et. al.,
J. Opt. B : Quantum Semiclass. Opt. {\bf 2}, 353358 (2000).

\bibitem{tkg} T. K. Ghosh, Phys. Lett. {\bf A 285}, 222 (2001),
cond-mat/0012188.

\bibitem{cz} D. Freedman and A. Newell (unpublished).

\bibitem{arnold} V. I. Arnold, Mathematical Methods of Classical
Mechanics, Second Edition ( Springer-Verlag, New York, 1989 ).

\bibitem{me1} Pijush K. Ghosh ( unpublished ).

\bibitem{kolo} E. B. Kolomeisky, T. J. Newman, J. P. Straley and X. Qi,
Phys. Rev. Lett. {\bf 85}, 1146 (2000).

\bibitem{mumu} R. K. Bhaduri, S. Ghosh, M. V. N. Murthy and D. Sen,
J. Phys. A: Math. Gen. {\bf 34}, 6553 (2001), cond-mat/0010075.

\bibitem{kagan} Yu. Kagan, E. L. Surkov and G. V. Shlyapnikov,
Phys. Rev. {\bf A 54}, R1753 (1996).

\bibitem{rp} L. P. Pitaevskii and A. Rosch, Phys. Rev. {\bf A 55},
R835 (1997), cond-mat/9608135.

\bibitem{rotate} A. L. Fetter and A. A. Svidzinsky, cond-mat/0102003.


\bibitem{jp} R. Jackiw and S.-Y. Pi, Phys. Rev. Lett. {\bf 64}, 2969 (1990);
{\bf 66}, 2682 (1991); Phys. Rev. {\bf D 42}, 3500 (1990); {\bf D 44}, 2524
(1991).

\bibitem{ez} Z. F. Ezawa, M. Hota and A. Iwazaki, Phys. Rev. Lett.
{\bf 67}, 411 (1991).

\bibitem{me} Pijush K. Ghosh, Phys. Rev. {\bf D 53}, 2248 (1996),
hep-th/9503199.
\end{thebibliography}
\end{document}